# Green and Gold Open Access Percentages and Growth, by Discipline

Yassine Gargouri,[1] Vincent Larivière,[3] Yves Gingras,[2] Les Carr[4] & Stevan Harnad[5,4]

[1] Institut des Sciences Cognitives, Université du Québec à Montréal
[2] Canada Research Chair in the History and Sociology of Science, Université du Québec à Montréal, Montréal (Canada)
[3] Observatoire des Sciences et des Technologies, Université du Québec à Montréal, Montréal (Canada)
[4] School of Electronics and Computer Science, University of Southampton, Southampton, United Kingdom
[5] Canada Research Chair in Cognitive Sciences, Université du Québec à Montréal, Canada

**Abstract**

Most refereed journal articles today are published in subscription journals, accessible only to subscribing institutions, hence losing considerable research impact. Making articles freely accessible online ("Open Access," OA) maximizes their impact. Articles can be made OA in two ways: by self-archiving them on the web ("Green OA") or by publishing them in OA journals ("Gold OA"). We compared the percent and growth rate of Green and Gold OA for 14 disciplines in two random samples of 1300 articles per discipline out of the 12,500 journals indexed by Thomson-Reuters-ISI using a robot that trawled the web for OA full-texts. We sampled in 2009 and 2011 for publication year ranges 1998-2006 and 2005-2010, respectively. Green OA (21.4%) exceeds Gold OA (2.4%) in proportion and growth rate in all but the biomedical disciplines, probably because it can be provided for all journals articles and does not require paying extra Gold OA publication fees. The spontaneous overall OA growth rate is still very slow (about 1% per year). If institutions make Green OA self-archiving mandatory, however, it triples percent Green OA as well as accelerating its growth rate.

**Introduction**

Peer-reviewed research in most disciplines today is published in subscription journals, accessible only to users at institutions that can afford to subscribe to the journal in which it is published. This means that research is losing the potential usage and impact of all users at institutions that cannot afford to subscribe to the journal in which it was published. In the print-on-paper era there was no remedy for this, but in the web era it has become possible for researchers to make their research freely accessible online to all would-be users ("Open Access," OA), not just subscribers, thereby maximizing research uptake and usage (Gargouri et al, 2010).

There are two ways researchers can make their articles OA: (1) by publishing in an OA journal – i.e., a journal that makes all of its articles freely accessible online (this is called "Gold OA") or (2) by publishing in any appropriate journal, but, in addition, "self-archiving" a supplementary copy free for all on the Web (usually on the author's institutional website). This is called "Green OA" (Harnad et al, 2004).

Green OA self-archiving has been possible since the advent of the Web in 1990, and even earlier, on the Internet (via "anonymous FTP"). Gold OA journals first began appearing around 1990, and their numbers have been growing ever since. There are now about 8000 Gold OA journals (as indexed by the Directory of Open Access Journals), which is about a third of all 25,000 peer-reviewed journals (as indexed by Ulrichsweb). However, among the most important peer-reviewed journals in most disciplines – the journals indexed by Thomson-Reuters-ISI – the proportion of Gold OA journals is much lower (in 2010 it was 625 out of the 12,500 ISI-indexed journals or about 5%) and the proportion of Gold OA journals among the top journals in each discipline is lower still. In addition, many Gold OA journals – and especially the top Gold OA journals – charge authors a publication fee, and sometimes a sizeable one.

In contrast to the constraints on providing Gold OA (quantity, quality and cost to the author) the only constraint on providing Green OA is whether or not the author actually chooses to self-archive. As a consequence, there is considerably more immediate scope for providing Green OA than for providing Gold OA. Björk et al (2010) estimated OA over all disciplines at 20.4% of all articles published (8.5% Gold OA, 11.9% Green) as sampled in 2009. In their sample, for the ISI-indexed journals, percent Gold was lower, (and lower than percent Green: 6.6% Gold, 14% Green OA) than in non ISI-indexed journals (14.2% Gold, 5.5% Green), with percent Green higher than Gold in all disciplines except the life sciences. We report here an update on these estimates for the ISI subset on a larger and more recent sample and range of articles, disciplines and years.

**Overall OA Percentages**

We estimated what percentage of the journal articles in 14 different scholarly and scientific disciplines published in the year range 2005-2010 were freely available on the Web (OA) in September 2011. A total sample of 107,052 articles was selected randomly from the Thomson-Reuters-ISI database so as to cover about 1,300 articles per year per discipline for each of the five years and 14 disciplines. Using each article's bibliographic metadata, a software robot then trawled the web and applied an algorithm to estimate which of the articles was available as OA full-text. The robot's accuracy, tested by hand checking a sub-sample of 200 articles, is about 98% (99% for OA articles and 98% for non-OA articles).

**Table 1** shows the average percent OA in 2011 for articles published in 2005-2010 for each year and discipline separately as well as averaged across all 14 disciplines. Percent OA varied from a low of 10% for Arts to a high of 45% for Mathematics (yearly average about 24% overall). **Figure 1** shows that for each discipline, its percent OA was about the same for each publication year within our range of 2005-2010.

**Table 1. Percent OA by discipline as measured in 2011 for articles published 2005-2010**

|  | All Discip | Math | Physics | Chemistry | Earth & Space | Engin. & Tech. | Bio. | Biom. Res. | Clin. Med. | Health | Psycho. | Social Sc. | Arts | Human. | Profess. Fields |
|---|---|---|---|---|---|---|---|---|---|---|---|---|---|---|---|
| Average 2005-2010 | 24% | 45% | 27% | 11% | 38% | 24% | 24% | 13% | 14% | 17% | 28% | 36% | 10% | 16% | 31% |
| 2005 | 24% | 43% | 26% | 13% | 39% | 28% | 26% | 13% | 17% | 18% | 29% | 38% | 9% | 14% | 31% |
| 2006 | 24% | 43% | 29% | 13% | 42% | 22% | 26% | 15% | 12% | 16% | 30% | 37% | 10% | 13% | 34% |
| 2007 | 23% | 46% | 23% | 9% | 41% | 24% | 23% | 15% | 12% | 16% | 30% | 36% | 8% | 13% | 33% |
| 2008 | 24% | 53% | 30% | 11% | 35% | 23% | 24% | 12% | 14% | 19% | 26% | 33% | 8% | 15% | 27% |
| 2009 | 23% | 42% | 29% | 10% | 32% | 23% | 24% | 13% | 13% | 15% | 29% | 36% | 13% | 20% | 29% |
| 2010 | 23% | 42% | 27% | 9% | 37% | 23% | 22% | 12% | 14% | 17% | 23% | 37% | 14% | 19% | 29% |

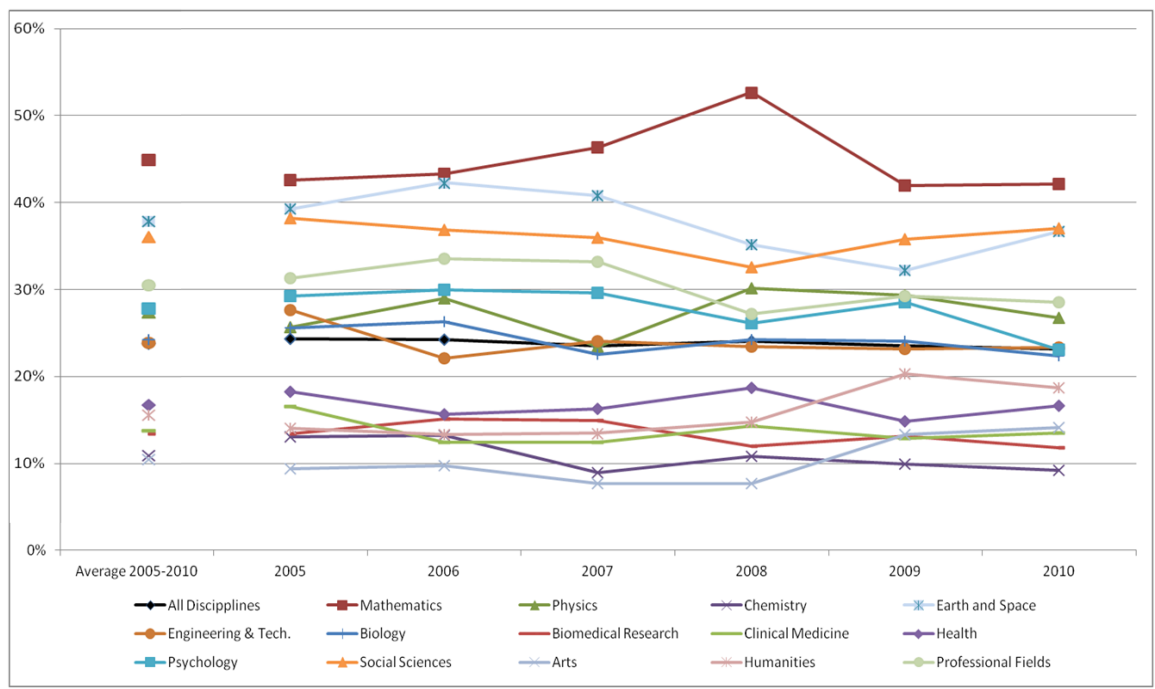

**Figure 1. Percent OA by discipline for articles published 2005 - 2010 (as measured in 2011)**

Using a similar method, we had made an earlier estimate in September 2009 on the basis of an ISI sample of 110,212 articles published 1998-2006 in 11 of the 14 disciplines (minus Arts, Humanities and Professional Fields[1]). **Table 2** shows the average percent OA for each year and discipline separately as well as averaged across all 11 disciplines.

---

[1] Professional fields includes a variety of professional disciplines related to Communication, Education, Information Science & Library Science, Law, Management, Miscellaneous Professional Fields and Social Work.

The overall average percent OA is about 20%, growing from 14% in 1998 to 21% in 2006 (**Figure 2**).

**Table 2. Percent OA by discipline for articles published 1998 - 2006 (as measured in 2009)**

|  | All Disci | Math | Phys | Chemis. | Earth Sc. | Engin. | Biology | Biomed. Research | Clinical Medicine | Health | Psycho | Social Sc. |
|---|---|---|---|---|---|---|---|---|---|---|---|---|
| **Average 1998-2006** | **20%** | 34% | 26% | 6% | 27% | 17% | 21% | 11% | 3% | 18% | 25% | 28% |
| **1998** | **14%** | 21% | 20% | 3% | 19% | 10% | 19% | 9% | 1% | 21% | 18% | 14% |
| **1999** | **16%** | 28% | 23% | 4% | 20% | 10% | 23% | 11% | 3% | 24% | 18% | 17% |
| **2000** | **18%** | 34% | 24% | 3% | 24% | 11% | 23% | 7% | 2% | 24% | 25% | 21% |
| **2001** | **20%** | 32% | 30% | 6% | 25% | 18% | 25% | 10% | 5% | 19% | 26% | 25% |
| **2002** | **20%** | 34% | 29% | 7% | 26% | 21% | 20% | 10% | 3% | 12% | 30% | 28% |
| **2003** | **22%** | 35% | 32% | 7% | 34% | 24% | 15% | 12% | 3% | 16% | 31% | 33% |
| **2004** | **23%** | 39% | 28% | 8% | 31% | 22% | 25% | 14% | 3% | 15% | 29% | 38% |
| **2005** | **22%** | 39% | 25% | 7% | 33% | 19% | 23% | 14% | 3% | 16% | 27% | 38% |
| **2006** | **21%** | 40% | 26% | 6% | 30% | 15% | 19% | 11% | 3% | 18% | 22% | 38% |

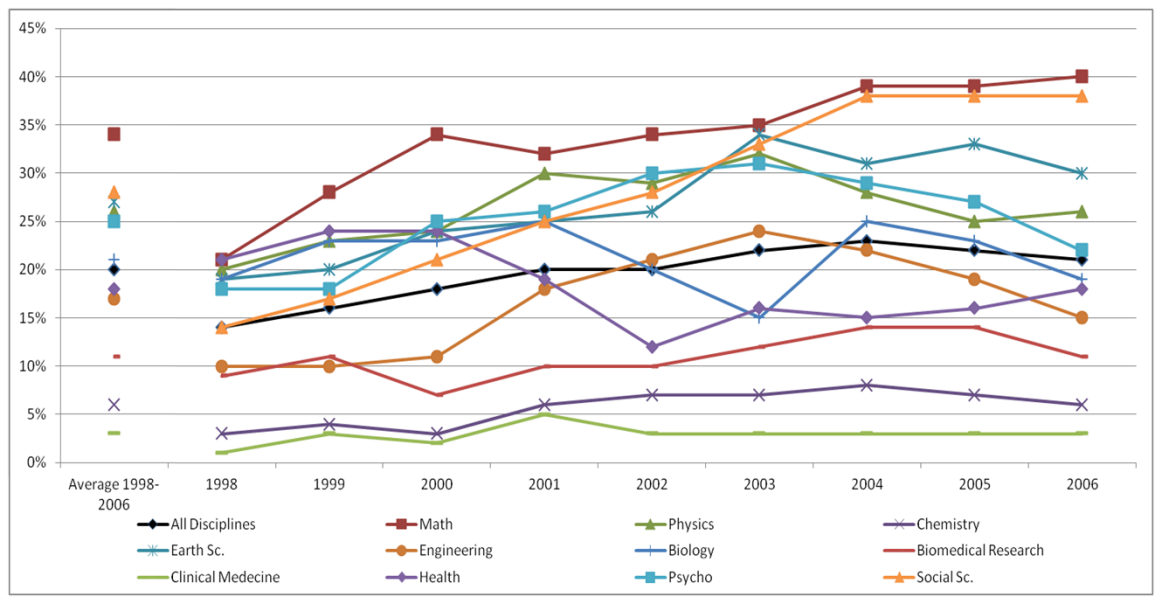

**Figure 2. Percent OA by discipline for articles published 1998 - 2006 (as measured in 2009)**

**Figure 3** compares percent OA for our 2009 sample (publication years 1998-2006) and our 2011 sample[2] (publication years 2005-2010) averaged across their respective year ranges. Percent OA has grown for all disciplines (especially Clinical Medicine (3% to 14%), Mathematics (34% to 45%), Chemistry (6% to 11%), Earth and Space Science (27% to 38%) and Social Science (28% to 36%).

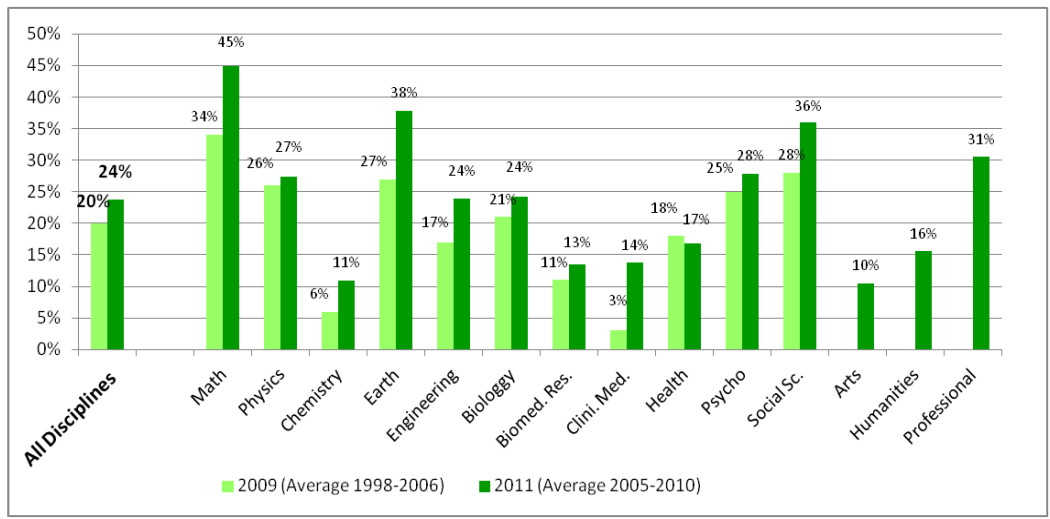

**Figure 3. Average percent OA (rounded to nearest percent) for the 2009 sample (for publication years 1998-2006) and 2011 sample (for publication years 2005-2010)**

A further comparison using only the two publication years that were covered by both samples (2005 and 2006) (**Figure 4)** shows a similar trend[3]: average percent OA grew for Clinical Medicine (3% to 14%), Mathematics (40% to 43%), Chemistry (7% to 13%), Earth and Space Science (32% to 41%) and Biology (21% to 26%) as well as overall. This growth is mainly explained by the retroactive self-archiving of researchers' past output.

---

[2] A paired sample t-test indicated that the mean growth of 5.5% (SD = 4.2, N=11) in percent OA from 2009 to 2011 was significantly greater than zero (t = 4.34, 2-tail p=0.001).

[3] A paired sample t-test indicated that the mean growth of 4.7% (SD = 3.8, N=11) in percent OA from 2009 to 2011 was significantly greater than zero (t = 4.05, 2-tail p = 0.002).

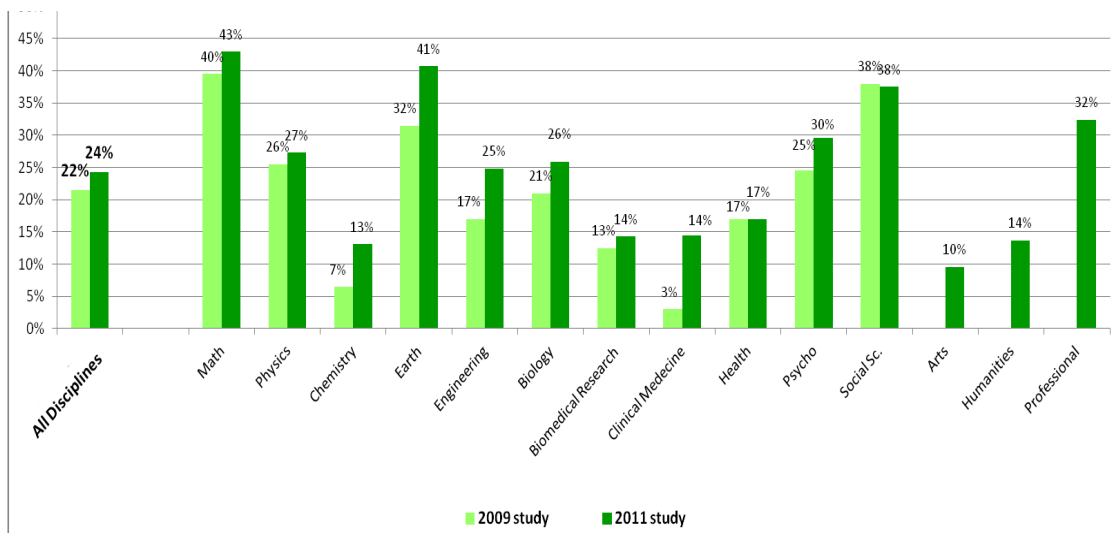

**Figure 4. Comparing average percent OA for the publication range covered in both samples (2005-2006) as measured in 2009 and then in 2011.**

**Separating Percent Gold and Green OA**

The robot-based estimates of percent OA were only used for the articles in our random sample that had been published in subscription journals (hence Green OA), but these were the vast majority. For the minority of articles in our random sample that had been published in (Gold) OA journals (as determined by whether they were indexed in the Directory of Open Access Journals, DOAJ) our estimate was based on the exact count of OA articles rather than on robot estimates. The percentage of the journals in our random sample that were OA journals was about 4% whereas the percentage of the articles published in OA journals was 2% (suggesting that OA journals publish fewer articles than subscription journals).

In our 2011 total random sample for articles published 2005-2010, about 97% of them were published in subscription journals, with 21% of them freely available on the web (Green OA). Adding the 2.4% of them that were published in OA journals (Gold OA) gave an overall percent OA of about 24% (**Figure 5**).

Social Science (0.9%), Chemistry (1.1%), Engineering and technology (1.3%) and Professional fields (1.3%) had the lowest percent Gold OA, whereas Biomedical Research (7.9%), Clinical Medicine (5.1%) and Health (4.6%) had the highest percent Gold OA (as in Bjork et al's 2010 study).

**Table 3. Percent Gold and Green OA (measured in 2011) for 2005-2010**

|  | All Discip | Math | Physics | Chem. | Earth | Engin. | Bio. | Biomed Res. | Clinical Medecine | Health | Psycho | Social Sc. | Arts | Human. | Professional Fields |
|---|---|---|---|---|---|---|---|---|---|---|---|---|---|---|---|
| **Green OA** | *21%* | *43%* | *26%* | *10%* | *36%* | *23%* | *22%* | *6%* | *9%* | *12%* | *27%* | *35%* | *9%* | *14%* | *29%* |
| ***Gold*** | *2%* | *2%* | *2%* | *1%* | *2.0%* | *1%* | *2%* | *8%* | *5%* | *5%* | *1%* | *1%* | *1%* | *1%* | *1%* |

| | | | | | | | | | | | | | | |
|---|---|---|---|---|---|---|---|---|---|---|---|---|---|---|
| *OA* | | | | | | | | | | | | | | |
| Overall OA | 24% | 45% | 28% | 11% | 38% | 24% | 24% | 14% | 14% | 17% | 28% | 36% | 10% | 15% | 30% |

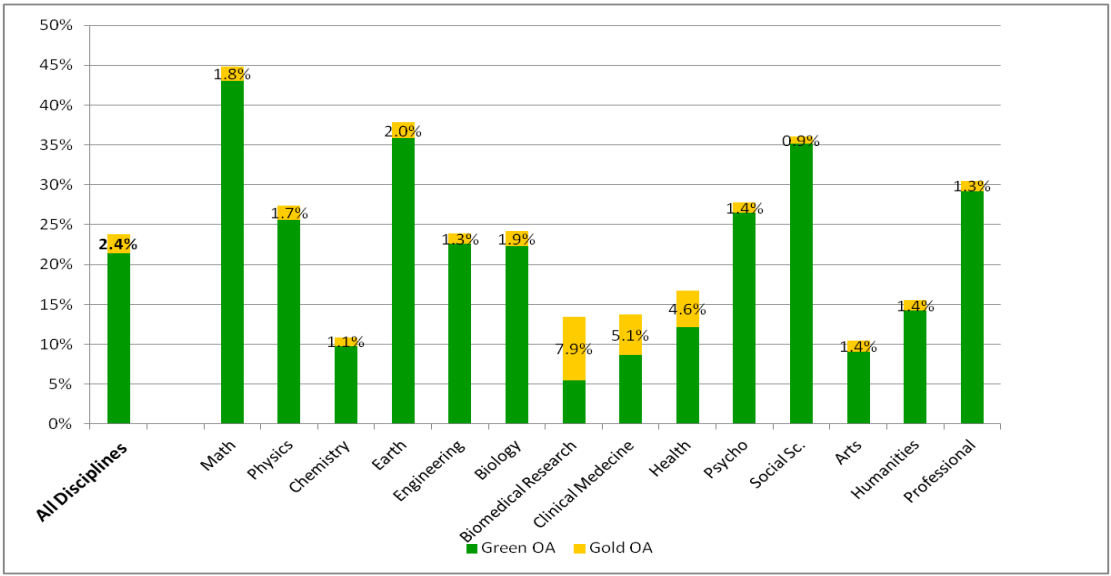

**Figure 5. Percent Gold and Green OA (measured in 2011) for 2005-2010**

**Table 4** and **Figure 6** show the percent Gold and Green OA for each publication year 2005-2010, for all disciplines as well as separately. Percent Gold for all disciplines remained relatively low (1.2% - 3.5%). The most substantial increase in Gold OA was for Biomedical Research; 2.1% in 2005 to 8.3% in 2009 (**Figure 7**). For all other disciplines, percent Green is higher than percent Gold OA. For example, in Mathematics, although overall percent OA was as high as 52.6% in 2008, percent Gold OA was only 4.4%. The increase in overall percent OA is hence due mostly to the increase in Green OA.

**Table 4. Percent Gold and Green OA (measured in 2011) by publication per year 2005-2010**

|  | Total 2005-2010 | | 2005 | | 2006 | | 2007 | | 2008 | | 2009 | | 2010 | |
|---|---|---|---|---|---|---|---|---|---|---|---|---|---|---|
|  | Gr OA | Go OA | Gr OA | Go OA | Gr OA | Go OA | Gr OA | Go OA | Gr OA | Go OA | Gr OA | Go OA | Gr OA | Go OA |
| All Discip. | 21.4% | 2.4% | 22.4% | 1.9% | 22.6% | 1.7% | 21.8% | 1.7% | 20.6% | 3.5% | 21.2% | 2.3% | 21.9% | 1.2% |
| Math | 43.0% | 1.8% | 41.0% | 1.6% | 42.8% | 0.5% | 44.5% | 1.8% | 48.2% | 4.4% | 40.5% | 1.4% | 40.8% | 1.3% |
| Physics | 25.6% | 1.7% | 24.3% | 1.3% | 27.6% | 1.4% | 21.8% | 1.6% | 27.4% | 2.7% | 27.1% | 2.3% | 25.7% | 1.1% |
| Chemistry | 9.8% | 1.1% | 12.4% | 0.6% | 12.1% | 1.2% | 8.1% | 0.8% | 7.6% | 3.2% | 9.2% | 0.8% | 9.3% | 0.0% |
| Earth | 35.9% | 2.0% | 37.6% | 1.6% | 40.5% | 1.8% | 38.9% | 1.9% | 33.2% | 2.0% | 29.3% | 2.8% | 35.0% | 1.7% |
| Engineering | 22.6% | 1.3% | 26.8% | 0.9% | 20.6% | 1.5% | 23.4% | 0.7% | 21.5% | 2.0% | 21.8% | 1.4% | 21.9% | 1.4% |
| Biology | 22.3% | 1.9% | 24.1% | 1.5% | 24.2% | 2.1% | 20.7% | 1.9% | 21.1% | 3.1% | 22.0% | 2.1% | 21.8% | 0.5% |
| Biom. Res. | 5.5% | 7.9% | 11.3% | 2.1% | 12.7% | 2.4% | 11.7% | 3.2% | 5.4% | 6.5% | 4.9% | 8.3% | 10.5% | 1.3% |
| Clin. Med. | 8.6% | 5.1% | 7.8% | 8.7% | 10.0% | 2.4% | 10.1% | 2.3% | 7.9% | 6.4% | 11.2% | 1.8% | 12.7% | 0.8% |
| Health | 12.2% | 4.6% | 15.1% | 3.1% | 11.5% | 4.2% | 11.7% | 4.6% | 12.0% | 6.8% | 10.6% | 4.3% | 12.3% | 4.3% |
| Psycho | 26.5% | 1.4% | 28.4% | 0.9% | 28.5% | 1.4% | 29.1% | 0.5% | 23.6% | 2.5% | 26.8% | 1.7% | 22.0% | 1.1% |
| Social Sc. | 35.1% | 0.9% | 37.5% | 0.7% | 36.4% | 0.4% | 35.5% | 0.5% | 30.2% | 2.3% | 35.3% | 0.4% | 36.1% | 0.9% |
| Arts | 9.0% | 1.4% | 8.9% | 0.5% | 8.2% | 1.6% | 5.7% | 2.0% | 6.4% | 1.3% | 11.5% | 1.8% | 12.8% | 1.3% |
| Humanities | 14.2% | 1.4% | 13.0% | 1.0% | 11.9% | 1.4% | 12.6% | 0.9% | 11.9% | 2.9% | 19.3% | 1.0% | 17.7% | 1.0% |
| Professional | 29.2% | 1.3% | 30.3% | 1.0% | 32.6% | 1.0% | 32.2% | 1.0% | 25.0% | 2.2% | 27.4% | 1.8% | 27.9% | 0.6% |

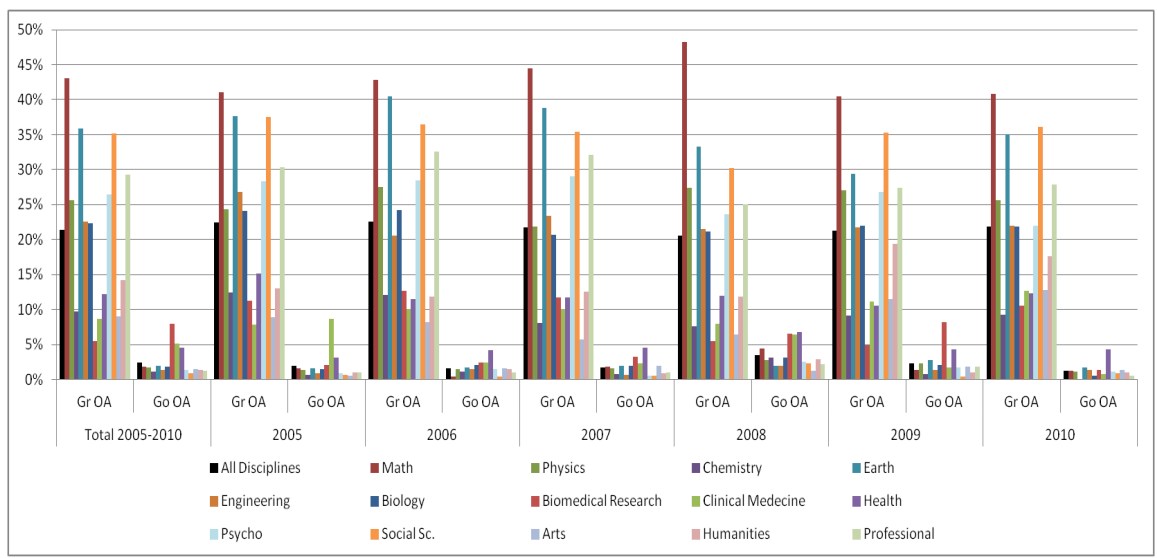

**Figure 6. Percent Gold and Green OA (measured in 2011) by publication per year 2005-2010**

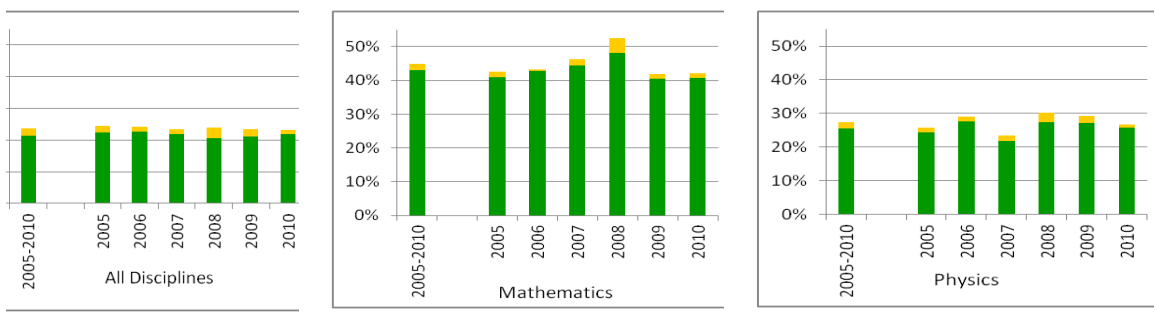

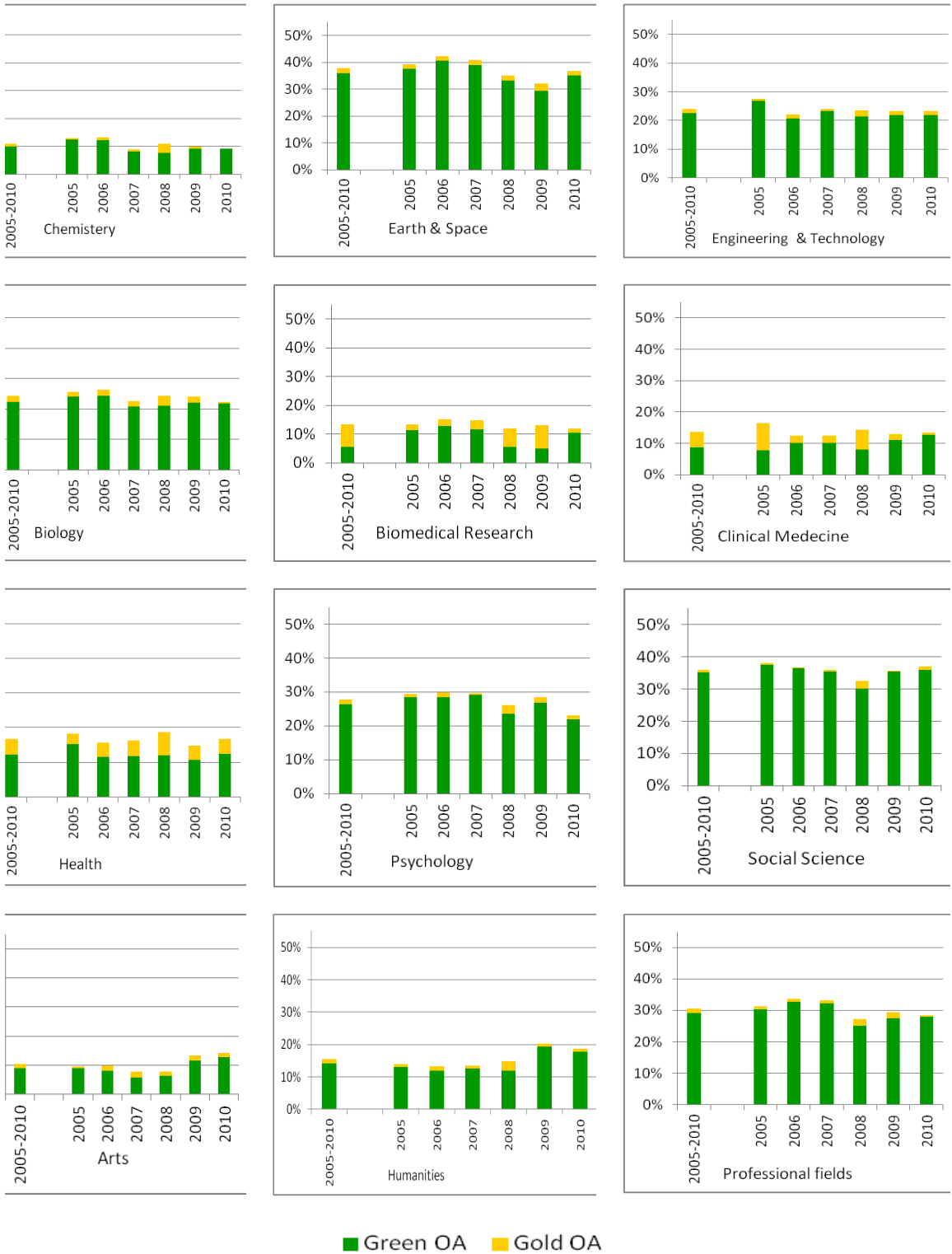

**Figure 7. Percent Gold and Green OA (measured in 2011) by discipline by publication per year 2005-2010**

**Conclusions**

Although Green OA self-archiving, with its much greater immediate scope for growth (many more subscription journals, many more of the top journals, no author fee) indeed exceeds Gold OA both in percentage and growth rate in almost all disciplines, neither its percentage nor its growth rate is anywhere near as great as it could be, if more authors self-archived. This underscores the fact that what is needed in order to maximize research access and impact is policies from researchers' institutions and funders mandating Green OA self-archiving (Harnad, 2011). Mandates almost immediately triple the baseline Green OA self-archiving rate (**Figure 8**), which continues climbing toward 100% OA in succeeding years thereafter.

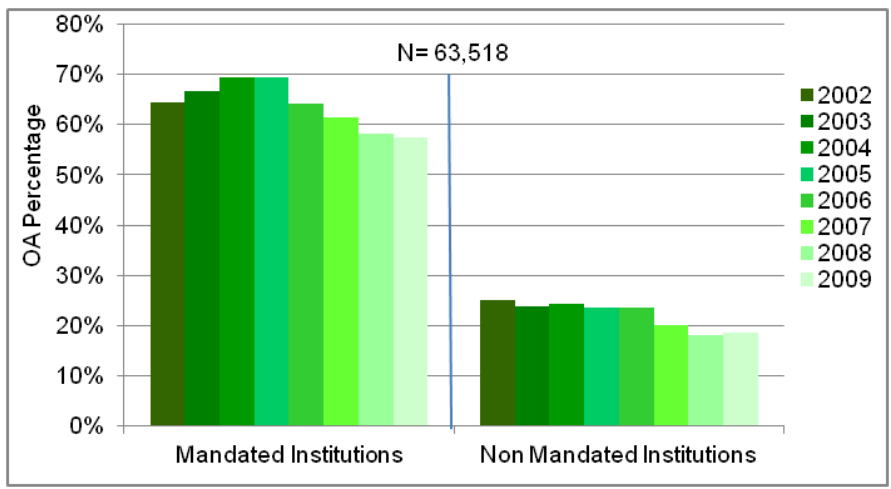

**Figure 8. Percent of research output that is green OA for institutions where Green OA is or is not mandatory (based on Gargouri et al's 2010 data, as reproduced from Poynder 2011).**

**Research Support:** SSHRC Grant 601077 "Open access to research: scientometric analysis of impact time-course" and Chaire de recherche du Canada en sciences cognitives (10950/581913)

**References**

Björk B-C, Welling P, Laakso M, Majlender P, Hedlund T, et al. (2010) Open Access to the Scientific Journal Literature: Situation 2009. *PLoS ONE* 5(6): e11273. doi:10.1371/journal.pone.0011273

Gargouri, Y., Hajjem, C., Lariviere, V., Gingras, Y., Brody, T., Carr, L. and Harnad, S. (2010) Self-Selected or Mandated, Open Access Increases Citation Impact for Higher Quality Research. *PLOS ONE* 5 (10) e13636 http://eprints.ecs.soton.ac.uk/18493/

Harnad, S. (2011) Open Access to Research: Changing Researcher Behavior Through University and Funder Mandates. *JEDEM Journal of Democracy and Open Government* 3 (1): 33-41. http://eprints.ecs.soton.ac.uk/22401/

Harnad, S., Brody, T., Vallieres, F., Carr, L., Hitchcock, S., Gingras, Y, Oppenheim, C., Stamerjohanns, H., & Hilf, E. (2004) The green and the gold roads to Open Access. *Nature Web Focus*. http://www.nature.com/nature/focus/accessdebate/21.html

Laakso M, Welling P, Bukvova H, Nyman L, Björk B-C, et al. (2011) The Development of Open Access Journal Publishing from 1993 to 2009. *PLoS ONE* 6(6): e20961. doi:10.1371/journal.pone.0020961

Poynder, Richard (2011) Open Access by Numbers. *Open and Shut*, 19 June 2011 http://poynder.blogspot.com/2011/06/open-access-by-numbers.html